# Dissociative electron attachment to HMX (octahydro-1,3,5,7-tetranitro-1,3,5,7-tetrazocine): Strong decomposition near 0 eV

Johannes Postler [1], Marcello M. Goulart [1,3], Carolina Matias [1], Andreas Mauracher [1], Filipe Ferreira da Silva [2], Paul Scheier [1], Paulo Limão-Vieira [2,*], Stephan Denifl [1,*]

[1] Institut für Ionenphysik und Angewandte Physik, Technikerstr. 25 / 3, A-6020 Innsbruck, Austria

[2] Laboratório de Colisões Atómicas e Moleculares, CEFITEC, Departamento de Física, Faculdade de Ciências e Tecnologia, Universidade Nova de Lisboa, 2829-516 Caparica, Portugal

[3] CAPES Foundation, Ministry of Education of Brazil, Brasilia – DF 70040-020, Brazil

**ABSTRACT**

Dissociative electron attachment (DEA) measurements to gas phase HMX, octahydro-1,3,5,7-tetranitro-1,3,5,7-tetrazocine, $C_4H_8N_8O_8$, have been performed by means of a crossed electron-molecular beam experiment. The most intense signals are observed at 46 and 176 u and assigned to $NO_2^-$ and $C_3H_6N_5O_4^-$, respectively. Anion efficiency curves for 15 negatively charged fragments have been measured in the electron energy region from about 0–20 eV with an energy resolution of ~ 0.7 eV. Product anions are observed mainly in the low energy region, near 0 eV arising from surprisingly complex reactions associated with multiple bond cleavages and structural and electronic rearrangement. The remarkable instability of HMX towards electron attachment with virtually zero kinetic energy reflects the highly explosive nature of this compound. The present results are compared to DEA results on the chemically related royal demolition explosive molecule (RDX).

[*] Corresponding authors. Fax: +43 512 507 2932; Fax: + 351 21 294 85 49
Email addresses: stephan.denifl@uibk.ac.at (S. Denifl); plimaovieira@fct.unl.pt (P. Limão-Vieira)

## 1. INTRODUCTION

The constant and actual need for detection of explosives and the development of techniques to distinguish them among several other similar yet unperilous substances has increased over the last years due to the high risk of terrorist attacks. One of the key issues is the fast capability to distinguish explosives amongst a background of other nitrogen-containing compounds. A large variety of mass spectrometric methods have been suggested for rapid explosive detection including laser photon ionization,[1,2] ion mobility spectrometry,[3,4] chemical ionization mass spectrometry,[5,6] and negative ion mass spectrometry based on free electron capture[7-10] (see also the extensive review by Moore[11]). Octahydro-1,3,5,7-tetranitro-1,3,5,7-tetrazocine, $C_4H_8N_8O_8$, commercially known as HMX, is a powerful and relatively insensitive nitroamine high explosive and chemically related to RDX - Royal Demolition Explosive (see Figure. 1). HMX and RDX are nitroamine polymers consisting of four and three $CH_2NNO_2$ units, respectively. Electrospray ionisation/ion mobility spectrometry[12] and Fourier transform ion cyclotron resonance mass spectrometry[13] was reported for elemental composition analysis of RDX and HMX, whereas plasma desorption mass spectrometry revealed $CN^-$ and $NO_2^-$ as the most abundant species.[14] To our best knowledge, no previous dissociative electron attachment study to HMX exists in the literature. Compared to other explosives HMX has a particular low vapour pressure[15] representing a challenging compound for experiments in laboratory as well as detection in the context of true-to-life situations.

During the last five years we have undertaken a series of studies on electron attachment to several (aromatic) nitro compounds used as explosives.[16-26] They have been performed in crossed electron-molecular beam experiments with high-energy resolution (~70 meV)[22] or high sensitivity (utilizing ~10 μA of electron current).[27] These experiments have included free electron interactions with bare molecules in the gas-phase[16-26] and embedded in He droplets.[28] In the former, and generally speaking, we have observed that capture of an excess electron with virtually no kinetic energy leads to formation of a variety of dissociative electron attachment (DEA) fragments reflecting therefore the explosive nature of the compounds, whereas in the latter the ultra-cold environment efficiently quenches all the gas-phase dissociation channels. DEA studies to explosives yielding $NO_2^-$ formation, allowed considering this fragment to serve as a fingerprint for the identification of the neutral compound. From these studies, some are related to the present molecule (HMX) with particular attention being paid to RDX.

In the present work we investigate the negative ion formation from HMX upon free electron capture at low electron energies (0 – 20 eV) by recording the ion yield as a function of the electron energy with a modest electron energy resolution of ~ 0.7 eV. By far the two most dominant signals are due to formation of $C_3H_6N_5O_4^-$ and $NO_2^-$. It is shown that at threshold (~ 0 eV) a variety of intense DEA products are formed.

## 2. EXPERIMENTAL AND COMPUTATIONAL DETAILS

Electron attachment to HMX was investigated by means of a crossed electron-molecular beam set-up utilizing a double focusing two-sector field mass spectrometer equipped with a standard Nier-type ion source.[27] The electron energy spread close to 0 eV is about 0.7 eV (FWHM), but the source is characterized by the high sensitivity and the rapid extraction of the anions from the ion source (less than 1μs). The electron current is regulated to 50 μA, which is reached for electron energies higher than 15 eV. Below that, the electron current varies linearly. A voltage drop of 6 kV is accelerating the ions from the ion source towards the sector fields. Negative ion yields are obtained as a function of the electron energy. HMX is solid at room temperature and therefore has to be heated in order to increase its vapour pressure so that at moderately elevated temperatures an effusive molecular beam can be generated. The effusive molecular beam emerges from a heated oven through an orifice of 3 mm diameter operated at a temperature of around 97° C which is well below the melting point of HMX. Thermal decomposition of HMX starts at temperatures above the melting point and reaches a maximum at 208 °C. The HMX sample was obtained from defusing section of the Austrian ministry of interior and contained an unknown amount of hydrocarbons. These impurities however led to no contamination of anion yields from HMX. The electron energy scale and the electron energy resolution are calibrated using the well-known $SF_6^-/SF_6$ signal near 0 eV and the resonances of the $F^-/SF_6$ and $F_2^-/SF_6$ anions at higher electron energies.[29]

To complement the experimental results, we have performed quantum chemical calculations utilizing Møller-Plesset perturbation theory truncated at the second order (MP2) for geometry optimizations, visualization of the molecular orbitals and calculation of energetics together with the 6-311++G(2d,p) basis set. The uncertainty of the energies derived at this level of theory and basis set is approximately at ± 0.20 eV and was derived from a series of test calculations for nitro-organic compounds, where we compared the energetics derived from MP2/6-311++G(2d,p) with values from G4(MP2) calculations, which have a known uncertainty of approximately ± 0.1 eV.[30]

## 3. RESULTS AND DISCUSSION
**A. Negative Ion Mass Spectrum**

Figure 2 (upper diagram) shows the negative ion mass spectrum of HMX obtained by summation of individual mass spectra measured at several different electron energies, i.e. from an energy close to 0 eV up to 10 eV in eight steps. From this figure it is possible to get an overview of all fragment anions formed, despite the fact that for a mass spectrum recorded at one electron energy only those anions appear which are produced in a resonance near the chosen electron energy. The most significant anions are listed in Table I together with the position of the corresponding resonances and the relative contribution to the total anion yield (integrated over the whole electron energy range studied).

In general, capture of a free electron by a polyatomic molecule (represented as ABC) generates a transient negative ion (TNI), (ABC)$^{\#-}$, that may further decompose via the following processes:

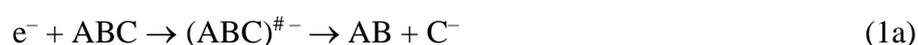
$$e^- + ABC \rightarrow (ABC)^{\#-} \rightarrow AB + C^- \qquad (1a)$$

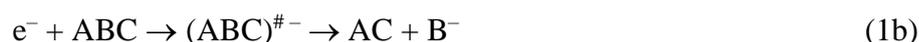
$$e^- + ABC \rightarrow (ABC)^{\#-} \rightarrow AC + B^- \qquad (1b)$$

The TNI is seen as a quasi-bound state embedded in the autodetachment continuum and unstable towards the loss of the extra charge. According to the negative ion mass spectra of HMX we measured, autodetachment or fragmentation occurs in a time window shorter than the detection time, resulting in the absence of an observable parent negative ion. This result is analogous to RDX previously studied in our laboratory[22] and is also reinforced by the considerable change in geometry from the neutral to the anion, resulting in the delocalization of the extra charge over one of the $NO_2$-groups (Figure 3), leading to fragmentation. Generally speaking, the formation of fragment ions in HMX through dissociative electron attachment (DEA) is most intense in features close to 0 eV.

The ion yields can be classified into four different groups according to their magnitudes: i) $C_3H_6N_5O_4^-$ fragment (176 u) and $NO_2^-$ fragment (46 u) are the most dominant anions; ii) the second group comprises $OH^-$ (17 u) and the species at 102 and 129 u where the former can be formed by a loss of an HCN from the anion at 129 u and the latter may result from the metastable parent anion $M^{\#-}$ via loss of $NO_2HNO_2HNO_2HCN$; iii) the third comprises the masses 16, 26, 28, 60, 82, 93, 156 and 160 u, which have been identified as $O^-$, $CN^-$, $CH_2N^-$, $NNO_2^-$, $[M-CNNO_2H_2NO_2H_2NO_2NO_2]^-$, $[NO_2HNO_2]^-$, $[M-3NO_2-2H]^-$ and $[C_2H_2N_5O_4]^-$, respectively; iv) masses 203 and 250 u are assigned as $[M-NO_2HNO_2]^-$ and $[M-NO_2]^-$, respectively. The bottom diagram in Figure 2 shows the mass spectrum of HMX obtained at 0 eV. The mass spectrum obtained close to 0 eV shows a rich chemistry driven upon low-energy electron attachment to HMX which was also found for RDX (see Figure 2 in Ref. 22) lending evidence to their more explosive character when compared to other explosives such as TNT[20] and the other nitrocompounds.[16-19,23,24,26]

In Figure 3 we show the fully optimized geometry obtained at MP2/6-311++G(2d,p) level of theory and basis set for the neutral molecule (upper left part) and the highest occupied molecular orbital (HOMO, MO 76) (upper right part) derived from the generalized density at MP2 level. Additionally, the optimized geometry of the anionic structure (lower left part) is shown. It can be seen that the geometry shows noticeable changes when an excess charge is added to the molecular system, with a clearly extended N–$NO_2$ bond (from 1.38 Å to 2.30 Å). The singly occupied molecular orbital (SOMO, MO 77) is also shown in Figure 3 (lower right part). The excess electron is localized in one of the four $NO_2$-groups in particular that from the extended N–$NO_2$ bond. Along the ring there are at least two nodes along the C–N bonds adjacent to the extended N–$NO_2$ bond, lowering therefore the stability of the ring structure. The adiabatic electron affinity of HMX is 1.35 eV (Table II).

**B. Ion yield curves**

The anion efficiency curves[33] of the most intense fragment anions observed in the negative ion mass spectra are shown in Figures 4 and 5. The most intense DEA signals can be found close to 0 eV. Several fragment anions are also formed in an extended electron energy range showing resonance features at around 2, 5 and 10 eV. However, the relative abundance of these high-energy resonances is for most of the fragment anions more than at least one order of magnitude lower than the feature close to 0 eV. For heavier fragment anion masses the low energy resonance is more dominant, i.e. these anions decay into lower mass fragments if they are formed at high electron impact energies. Four out of 15 fragments are not predominantly formed close to 0 eV. These fragments were identified to be $O^-$, $CN^-$, $CH_2N^-$ and $NNO_2^-$ at 16, 26, 28 and 60 u, respectively. While the TNIs generated at low energies may be assigned as shape resonances involving the π* antibonding orbitals, it is likely that the resonance features at higher energies can be characterized as core excited resonances with possible contributions of high-energy shape resonances.

Assuming the general case of a polyatomic molecule (ABC), the threshold energy ($E_{th}$) of the DEA reaction (1a) is given by (regarding energy conservation):

$$E_{th} = D(AB - C) - EA(C) \qquad (2a)$$

with $D(AB-C)$ the dissociation energy and $EA(C)$ the electron affinity of the (neutral) fragment carrying the extra charge after the capture event. As far as standard heats of formation ($\Delta H_f^0$) are concerned, Eq. (2a) can be written as:

$$E_{th} = \Delta H_R^0 = \Delta H_f^0(AB) + \Delta H_f^0(C^-) - \Delta H_f^0(ABC) \qquad (2b)$$

where $\Delta H_R^0$ stems for the standard reaction enthalpy of reaction (1a), and $\Delta H_f^0(C^-) = \Delta H_f^0(C) + EA(C)$. From the energy balance, the threshold energy for reaction (1a) is typically below 4 eV if, generally speaking, the negative ion is formed by simple bond-cleavages and no rearrangement processes in a neutral fragment take place. This is due to the fact that the electron affinity for most radicals is below the bond dissociation energy. However, for more complicated reactions involving rearrangement, as is represented in reaction (1b), the energy gain by formation of the highly stable neutral products AC can shift the threshold energy to very low values.

*1. $NO_2^-$ (46 u), $[M–NO_2]^-$ (250 u) and $[M–NO_2HNO_2HCN]^-$ (176 u)*

The anions *$[M–NO_2]^-$ and $NO_2^-$* are formed via the cleavage of one of the four N–N bonds leading to the complementary DEA reactions with respect to the extra charge:

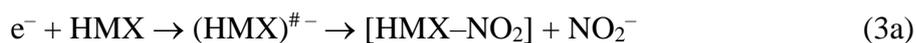

$$e^- + HMX \rightarrow (HMX)^{\#-} \rightarrow [HMX-NO_2] + NO_2^- \qquad (3a)$$

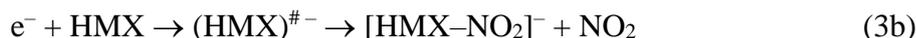

$$e^- + HMX \rightarrow (HMX)^{\#-} \rightarrow [HMX-NO_2]^- + NO_2 \qquad (3b)$$

Figure 4a shows the ion yield curve for $[HMX-NO_2]^-/C_4H_8N_7O_6^-$ and Figure 4b for $NO_2^-$. We derived the energetic thresholds for reactions (3a) and (3b) at MP2/6-311++G(2d,p) level of theory and basis set to be endothermic for (3a) by 0.48 eV and exothermic for (3b) by 0.91 eV. One has to consider that these values do not describe the actual path along the corresponding reaction coordinate but the final states. The medium- and high-energy features at about 5.3 and 10 eV of the anion yield curve of $NO_2^-$ (Table I) are most likely due to core-excited resonances. A quick look at Figure 4 reveals that the signal of $NO_2^-$ extends to higher energies, in contrast to its complementary fragment anion $[M–NO_2]^-$ restricted to the ~0 eV resonance only. If we assume that the excess energy of the transient negative ion $M^{\#-}$ is statistically distributed into the vibrational degrees of freedom, the large fragment should carry away about 95%. As far as $NO_2^-$ detection is concerned, the neutral counterpart will become increasingly unstable towards dissociation which does not influence the anion efficiency curve of $NO_2^-$. However, if the charge localizes on the heavy fragment, the excess energy will drive further decomposition into lower mass fragments and thus suppressing the anion yield for 250 u at higher electron energies. A careful analysis of the fragment anion yields in Figures 4 and 5 with masses lower than 250 u, shows remarkably that heavier fragment anions (between 129 and 203 u) show the resonance at medium energies with suppression of the high energy feature while all lighter fragment anions ≤ 102 u exhibit the high energy resonance as well. This indicates that the subsequent decomposition of [M–$NO_2$]$^{\#-}$ may contribute to the formation of these anions.

The 176 u anion efficiency curve shown in Figure 4e is assigned to [M–$NO_2$HNO$_2$HCN]$^-$, i.e. $C_3H_6N_5O_4^-$. Another possible fragment anion with mass 176 u would be the ammonium 5-nitrotetrazolate anion, $C_2H_4N_6O_4^-$. However, the detailed analysis of the isotope pattern leads to us to the assignment of $C_3H_6N_5O_4^-$. The resonance close to 0 eV indicates that no activation energy is required for this anion. The high energy resonances of this anion is barely discernible at ~4.7 eV.

*2. The complementary ions [M–NO$_2$HNO$_2$]$^-$ / $C_4H_7N_6O_4^-$ (203 u) and [NO$_2$HNO$_2$]$^-$ (93 u)*

The signal at 203 u (Figure 4c) can be identified as the ion arising from the loss of neutral $NO_2$ and $HNO_2$ (nitrous acid) units. The loss of such neutral units has been recently reported for RDX.[22] With the stoichiometric composition of the corresponding anion $C_4H_7N_6O_4^-$, such a reaction would require rearrangement including hydrogen transfer. The resonance profile of this anion shows the main feature close to zero energy and a weak resonance at 4.4 eV. It is particularly interesting to note that the losses of a $NO_2$ radical from the TNI and $HNO_2$ radical from [M–$NO_2$]$^-$ can actually be triggered by an excess electron at just 0 eV. With the thermochemical data of Table 2 for the reaction:

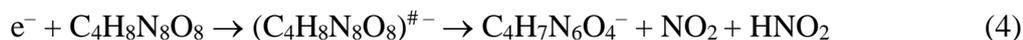
$$e^- + C_4H_8N_8O_8 \rightarrow (C_4H_8N_8O_8)^{\#-} \rightarrow C_4H_7N_6O_4^- + NO_2 + HNO_2 \qquad (4)$$

the $\Delta H_f^o$ ($C_4H_7N_6O_4^-$) ≥ 2.1 eV, where the exact value holds for the case when reaction (4) proceeds without excess energy, i.e., at the threshold energy).

We assign the complementary ion, 93 u, to [NO$_2$HNO$_2$]$^-$ (see Figure 4d) showing a resonance profile similar to the [M–NO$_2$HNO$_2$]$^-$ ion despite its intensity being higher by almost an order of magnitude. The dissociative electron attachment studies of RDX from Sulzer et al.,[22] have reported the complementary ions [RDX–$NO_2$–$HNO_2$]$^-$ and [NO$_2$HNO$_2$]$^-$, both formed within a narrow resonance close to 0 eV. They have observed that the extra electron possesses a stronger tendency to get localised on the heavier complement than on $NO_2HNO_2$, which is in clear contrast to HMX. Though, the only reason for such difference may reside on the fact that the [M–$NO_2HNO_2$]⁻ ion is more unstable towards further dissociation in HMX than in RDX.

*3. The anions $C_2H_2N_5O_4^-$ (160 u), [M–$NO_2HNO_2HNO_2$]⁻ (156 u) and [M–$NO_2HNO_2HNO_2HCN$]⁻ (129 u)*

The 160 u anion in Figure 4f shows a particularly strong low energy resonance close to 0 eV and a second low intensity feature at 4.4 eV. This fragment anion is tentatively assigned to $C_2H_2N_5O_4^-$, which results from the cleavage of several bonds and a series of intramolecular rearrangements. The 156 and 129 u (Figures 4g and 4h respectively) are almost exclusively formed via the low-energy resonance at ~0 eV, whereas the contribution at 4.5 eV is about two orders of magnitude lower. The only possible composition of these fragments requires the loss of three nitro groups together with two hydrogen atoms for the former and three hydrogen atoms and a CN for the latter.

*4. The anions $C_2H_4N_3O_2^-$ (102 u) and [M–$NO_2HNO_2HNO_2HNO_2HCN$]⁻ (82 u)*

With the exception of the $NO_2^-$ ion discussed above, the 102 and 82 u anions (including those discussed further below) are formed via at least three resonances as listed in Table I. The appearance of a fragment ion with 102 u has been reported in plasma desorption mass spectrometry of HMX by Hakansson *et al.*,[14] who tentatively assigned it to $C_4H_6O_3^-$. Further to recent DEA experiments to RDX[20] where a 102 u anion was detected and due to the structural similarity with HMX, we assign the fragment anion at 102 u rather to $C_2H_4N_3O_2^-$. In the present experiment we observe for this anion the third strongest ionic yield at ~0 eV. Figure 5a, shows the ionic yield for this anion which is formed via the following reaction:

$$e^- + C_4H_8N_8O_8 \rightarrow (C_4H_8N_8O_8)^{\#-} \rightarrow C_2H_4N_3O_2^- + NO_2 + C_2H_4N_4O_4 \quad (5)$$

With the thermochemical data of Table II, $\Delta H_f^o(C_4H_8N_8O_8) = 1.65$ eV, $\Delta H_f^o(C_2H_4N_4O_4) = -1.39$ eV and 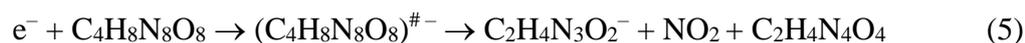 $\Delta H_f^o(NO_2) = 0.34$ eV, we get $\Delta H_f^o(C_2H_4N_3O_2^-) > 2.69$ eV. However, the observed low energy resonance close to 0 eV may indicate that the precursor anion is formed in a vibrationally excited state prior to dissociation. The resonances at 4.8 and 9.8 eV (Table

I) can also be assigned to electronically excited TNI states (including Rydberg excitation), which may decompose via one fragment anion plus one or more neutrals.

The 82 u anion in Figure 5b shows a low energy resonance close to 0 eV followed by another at 5.0 and 9.8 eV. The formation of this anion may involve the cleavage of several bonds and a series of intramolecular rearrangements. The resulting chemical composition of this fragment anion may be $C_3H_4N_3^-$.

*5. Other anionic yields: $NNO_2^-$ (60 u), $CH_2N^-$ (28 u), $CN^-$ (26 u), $OH^-$ (17 u) and $O^-$ (16 u)*

In Figure 5c and 5d the relative cross-sections for $NNO_2^-$, $CH_2N^-$ are shown, where these two anions are mostly formed through two resonances at around 5.5 and 9.8 eV with a relative ratio of 2:1 and 2.5:1, respectively. The 60 and 28 u anions show a weak contribution at low electron energies amounting only 6 and 3% of the strongest resonance. The DEA yield for the 26 u fragment can arise from the isobaric fragments $CN^-$ and/or $C_2H_2^-$. However, in previous experiments with RDX[22] and DNB and its deuterated analogues,[17] the formation of the vinyldene anion $(CH_2=C)^-$ in a complex rearrangement reaction was excluded. In the light of these findings we also assign the present signal at 26 u to the cyanide anion $CN^-$ (see Figure 5e for the anion efficiency curve and Table I for the estimated positions). The cyanide anion may be formed either by the excision of this unit from the target molecule or a complex reaction pathway via an intermediate anion. Since the cyano radical has an appreciable electron affinity (3.862 eV, Table II) which exceeds even that of the halogen atoms, its formation via complex DEA reactions has been reported and is well-known for amino acids and other large molecules containing C and N.[34-36] The resonant features for $CN^-$ in DEA to HMX can, as possible with $NO_2^-$, be compared with the other nitrocompounds as a means to distinguish chemical compounds with similar structures.

Figures 5f and 5g show the yields of the fragment anions found at 17 and 16 u as a function of the electron energy, respectively. These are assigned to $OH^-$ and $O^-$, with the former showing a considerable strong resonance close to 0 eV electron energy and two weak contributions at 4.6 and 9.7 eV (Table I). Such anion formation requires breaking an N=O and C–H bonds followed by rearrangement, where such concerted mechanism is remarkable at low incident electron energies. As far as $O^-$ is concerned, the high energy resonances between 6 and 12 eV are attributed to a contribution from the background signal from water, whilst the lowest resonance at about 2.2 eV is due to background signal from an unknown contamination present in the vacuum chamber.

**C. Detection of HMX vs. RDX and other nitroaromatic compounds**

Figure 7 shows a comparison of the anion efficiency curves of $NO_2^-$ as well as 102 and 60 u formed upon DEA to RDX plotted against HMX. An inspection of the figure reveals that the resonance positions for these common fragments anions are very similar in RDX and HMX. Thus a determination of HMX vs. RDX based on the resonance positions of common fragment anions like $NO_2^-$ seems not possible in an analytical application utilizing DEA. This problem arises due to the polymeric relation of these molecules, which for example also leads to a very similar electron ionization mass spectrum at 70 eV for these molecules.[31] In contrast, comparing with other explosives like TNT and non-explosive compounds like Musk Ketone a measurement of the corresponding anion yields (like for example $NO_2^-$ also included in Figure 6) may be well used as a fingerprint for the detection of these chemical compounds via the determination of resonance positions of common fragment anions. A detection of HMX vs. RDX in a DEA based application may be nevertheless also possible by measuring the anion yield of common fragment anions and subsequent determination of the ratio of resonance at different electron energies. For example, $C_2H_4N_3O_2^-$ of RDX shows a significant lower intensity of the zero eV contribution relative to the resonances located above ~0eV (see Fig. 7). Taking also into account the different intensity ratios of common anions, e.g. $C_3H_6N_5O_4^-/NO_2^-$ or $[M–NO_2HNO_2]^-/[NO_2HNO_2]^-$ discussed above, a differentiation of these explosives is possible. The latter is also important in view of possible health risks due to a supposed considerably higher toxicity and carcinogenicity of RDX compared to HMX.[37]

**4. CONCLUSION**

In the present study we have investigated DEA to HMX and determined the partial cross-sections of fragment anions measured in the electron energy range between ~0 eV and 20 eV with an energy resolution of 0.7 eV. The setup used is a commercial sector field mass spectrometer equipped with a Nier-type ion source which enables the detection of additional low-intensity fragments. Capture of an excess electron with virtually no kinetic energy by HMX leads to the formation of a variety of DEA fragments produced in a resonance near zero eV. The most dominant signal in DEA reactions to HMX is the formation of $C_3H_6N_5O_4^-$ (176 u) at electron energies below 5 eV whereas the second most intense signal assigned to $NO_2^-$ shows also another contribution at about 10 eV. Other and more complex reactions like the loss of several other neutral units with an onset of the resonant ionic yields at zero energy are observed as well. In comparison with other aromatic nitrocompounds, the absence of the non-decomposed anion and the rich and intense fragmentation already at electron energies close to 0 eV, confers the explosive character of HMX.

Previously it was proposed that the low-energy electron attachment resonance profiles obtained in DEA experiments can be used as unique characteristics for every molecule, making this sort of powerful experiments attuned to identifying small traces of chemically similar compounds. Separation of the polymers RDX and HMX seems not to be straight-forward by the measurements of the resonance profiles but possible by determining the intensity ratios of highly abundant anions, e.g. $C_3H_6N_5O_4^-/NO_2^-$. In this case DEA coupled with mass spectrometry can be used to act as a fingerprint in sensing and field explosive detection instrumentation.

## 5. ACKNOWLEDGEMENTS

This work has been supported by the Fonds zur Förderung der wissenschaftlichen Forschung (FWF), Wien, the European Commission, Brussels, via COST Action CM0805 programme "The Chemical Cosmos". PL-V and FFS acknowledge the PEst-OE/FIS/UI0068/2011 grant. M. M. G. acknowledges the National Council for the Improvement of Higher Education (CAPES), process nº 4752/11-2, the Foundation for Research Support of Minas Gerais State (FAPEMIG) and the National Council for Scientific and Technological Development (CNPq). We gratefully acknowledge the defusing section of the ministry of interior that provided us with HMX samples.

Figure 1. Chemical structures of HMX and RDX.

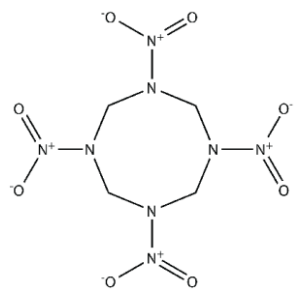 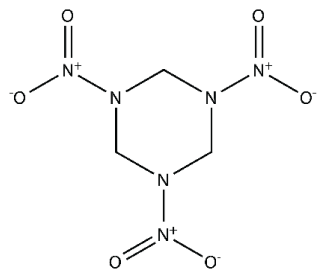

**HMX**            **RDX**

Figure 2. Negative ion mass spectrum of HMX obtained by summation of individual mass spectra at several different electron energies, i.e. from the electron energy close to 0 eV up to 10 eV in 8 steps (upper panel). The mass spectrum in the lower panel is measured at the electron energy close to 0 eV. Please note that both mass spectra show peaks at 127 u / 129 u ($SF_5^-$) and 146 / 148 ($SF_6^-$) formed by electron attachment to the calibration gas $SF_6$. However, the anion yield at 129 u originates both from the sample and the isotope of $SF_5^-$.

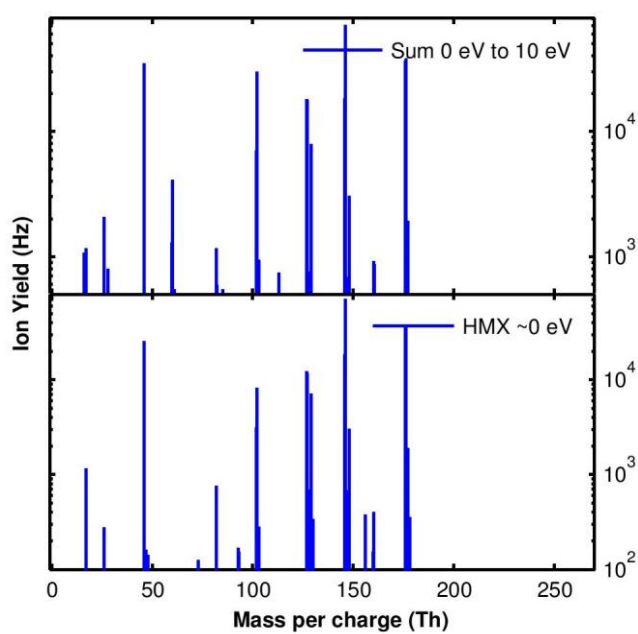

Figure 3. Optimized structures of HMX as neutral (upper left panel) and negatively charged (lower left panel) molecule. Highest occupied molecular orbital of the neutral molecule (upper right panel) and singly occupied molecular orbital of the anion (lower right panel). All results obtained at MP2/6-311++G(2d,p).

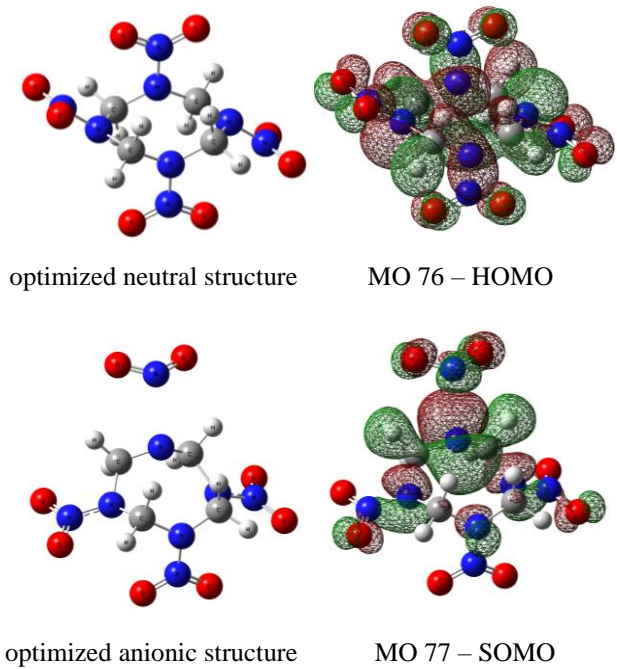

optimized neutral structure      MO 76 – HOMO

optimized anionic structure      MO 77 – SOMO

Figure 4. Anion efficiency curves of selected anions measured with a commercial sector field mass spectrometer equipped with a Nier-type ion source. The width of the electron energy distribution is about 1 eV and the electron current was set to 50μA.

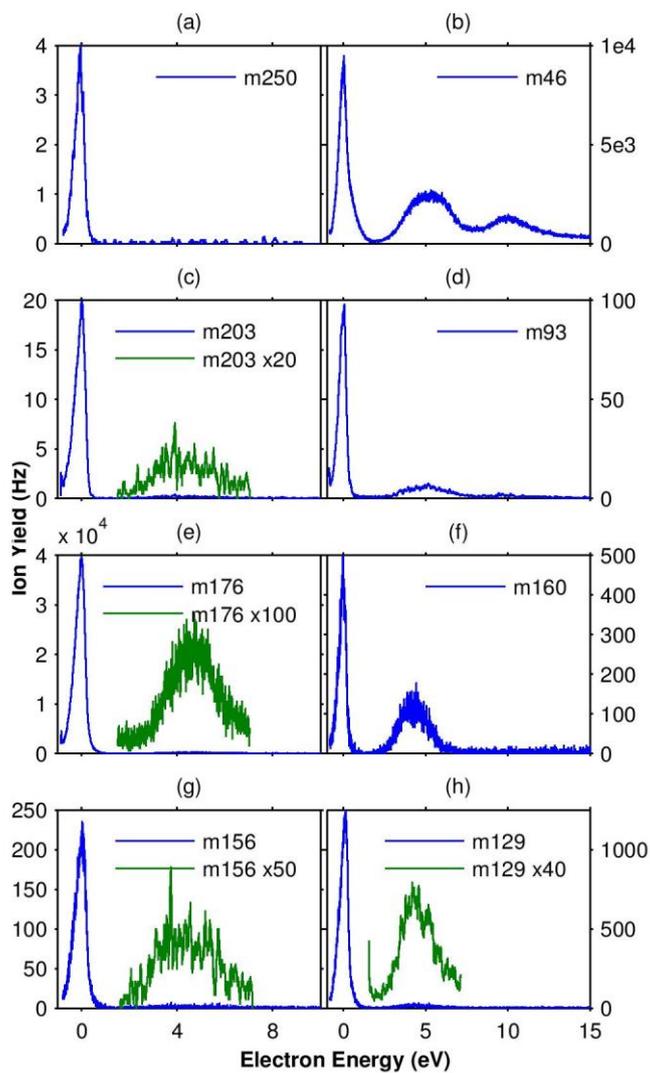

Figure 5. Anion efficiency curves of selected anions measured with a commercial sector field mass spectrometer equipped with a Nier-type ion source. The width of the electron energy distribution is about 1 eV and the electron current was set to 50μA.

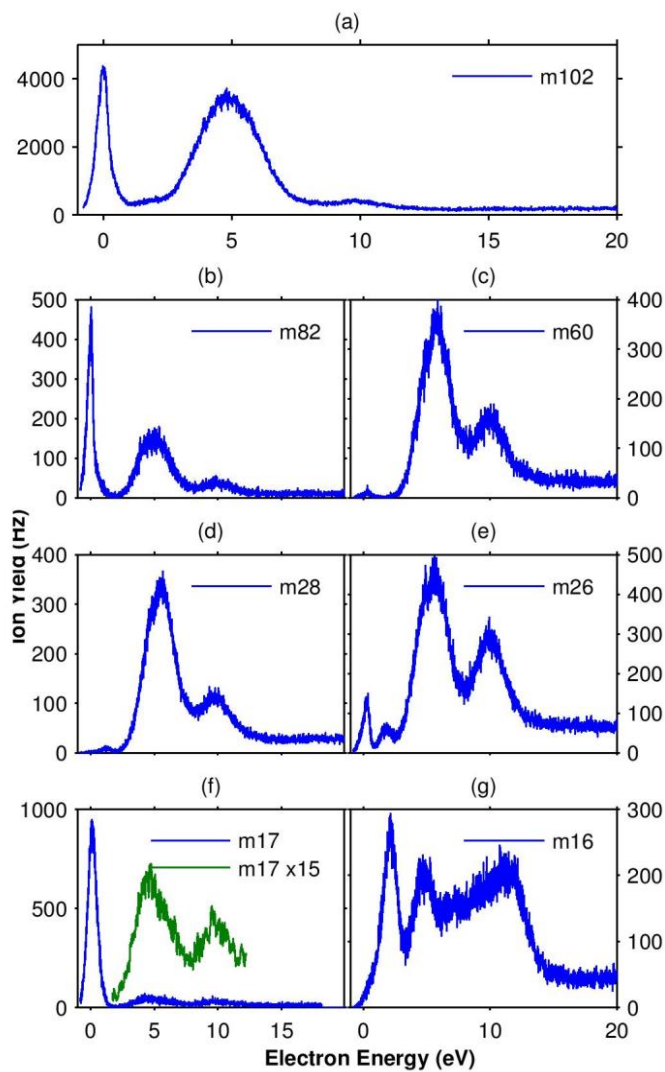

Figure 6. Ion yield curves for 102, 60 and 42 u from HMX and RDX. Note that the 0 eV peak in the RDX measurements is less broad due to the better electron energy resolution of the instrument used for the RDX measurements. [11]

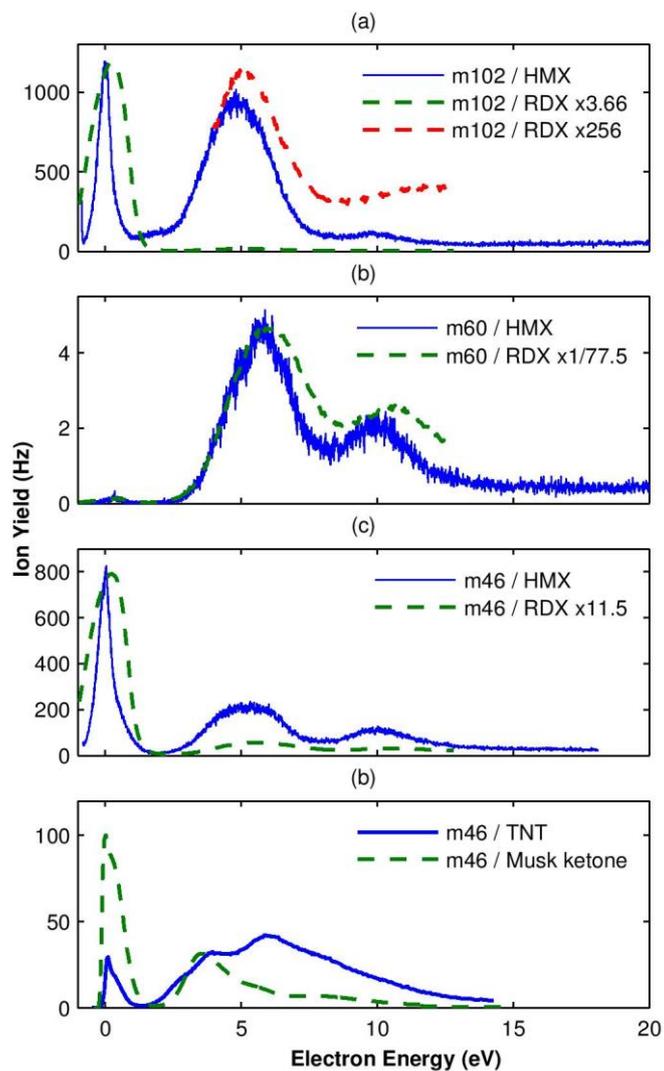

Table I – Peak positions for the fragment ions of HMX obtained in the present experiment.

| Mass (u) | Anionic species assignment | Relative intensity (%) | Peak position (eV) | | | |
|---|---|---|---|---|---|---|
| 250 | [M–NO$_2$]$^-$ | 0.002 | ~0 | … | … | … |
| 203 | [M–NO$_2$HNO$_2$]$^-$ | 0.013 | ~0 | … | 4.4 | … |
| 176 | C$_3$H$_6$N$_5$O$_4^-$ | 24.5 | ~0 | … | 4.7 | … |
| 160 | C$_2$H$_2$N$_5$O$_4^-$ | 0.83 | ~0 | … | 4.4 | … |
| 156 | [M–NO$_2$HNO$_2$HNO$_2$]$^-$ | 0.14 | ~0 | … | 4.4 | … |
| 129 | [M–NO$_2$HNO$_2$HNO$_2$HCN]$^-$ | 0.81 | 0.1 | … | 4.5 | … |
| 102 | CH$_2$NCH$_2$NNO$_2^-$ | 24 | ~0 | 1.9 [s] | 4.8 | 9.8 |
| 93 | NO$_2$HNO$_2^-$ | 0.13 | ~0 | … | 5.0 | 9.7 |
| 82 | C$_3$H$_4$N$_3^-$ | 1.51 | ~0 | … | 5.0 | 9.8 |
| 60 | NNO$_2^-$ | 3.83 | 0.3 | … | 5.8 | 9.8 |
| 46 | NO$_2^-$ | 34.1 | ~0 | … | 5.3 | 10 |
| 28 | CH$_2$N$^-$ | 3.05 | … | 1.2 | 5.5 | 9.8 |
| 26 | CN$^-$ | 6.16 | 0.2 | 1.8 | 5.4 | 9.9 |
| 17 | OH$^-$ | 1.18 | 0.1 | … | 4.6 | 9.7 |
| 16 | O$^-$ | 2.58 | … | 2.2 | 4.9 | … |

[s] means shoulder structure

Table II – Gas phase standard heats of formation ($\Delta H_f^\circ$) and electron affinities relevant in dissociative electron attachment to HMX (taken from Ref. 31, otherwise presently calculated value).

| Compound | $\Delta H_f^\circ$ (kJ mol$^{-1}$) |
|---|---|
| $C_4H_8N_8O_8$ (octahydro-1,3,5,7-tetranitro-1,3,5,7-tetrazocine, HMX) | 159 |
| $C_2H_4N_4O_4$ (1,1-diamino-2,2-dinitroethylene) | –133.9 [a] |
| $NO_2$ | 33.1 |
| $HNO_2$ (nitrous acid) | –76.73 |
| $NO_2HNO_2^-$ | – 400.0 ± 4.2 |
| $NO_2^-$ | 82.84 |

| Compound | Electron Affinity (eV) |
|---|---|
| HMX (calculated) | 1.35 |
| HMX–$NO_2$ (calculated) | 3.41 |
| $NO_2$ | 2.27 |
| CN | 3.862 ± 0.005 |

[a] from Ref. 32